%% file: 2014.CloudCom.WT.tex
\documentclass[10pt, conference, compsocconf]{IEEEtran}

\usepackage{amsmath}
\usepackage{listings}
\usepackage{graphicx}
\usepackage{courier}
\usepackage{float}
\usepackage{subfig}
\usepackage{tabularx}
\usepackage{pdfcomment}

\newcommand\textt[1]{{\scriptsize\textt{#1}}}

\newcommand{\essence}[0]{\textsc{Essence}}
\newcommand{\conjure}[0]{\textsc{Conjure}}

\begin{document}
%
\title{Optimal Deployment of Geographically Distributed\\Workflow Engines on the Cloud}

\author{\IEEEauthorblockN{Long Thai, Adam Barker, Blesson Varghese, Ozgur Akgun and Ian Miguel }
\IEEEauthorblockA{School of Computer Science, University of St Andrews, Fife, UK\\
Email: \{ltt2, adam.barker, varghese, ozgur.akgun, ijm \}@st-andrews.ac.uk}
}

\maketitle

\begin{abstract}
When orchestrating Web service workflows, the geographical placement of the orchestration engine(s) can greatly affect workflow performance. Data may have to be transferred across long geographical distances, which in turn increases execution time and degrades the overall performance of a workflow. In this paper, we present a  framework that, given a DAG-based workflow specification, computes the optimal Amazon EC2 cloud regions to deploy the orchestration engines and execute a workflow. The framework incorporates a constraint model that solves the workflow deployment problem, which is generated using an automated constraint modelling system. The feasibility of the framework is evaluated by executing different sample workflows representative of scientific workloads. The experimental results indicate that the framework reduces the workflow execution time and provides a speed up of 1.3x-2.5x over centralised approaches.  


\end{abstract}

\IEEEpeerreviewmaketitle

\section{Introduction}

\input{Intro}

\section{Workflow Deployment Problem}

\input{Model}

\section{Framework}

\input{Framework}

\section{Experimental Studies}
\input{Experiment}

\section{Related Work}

\input{LiteratureReview}

\section{Conclusion}

\input{Conclusion}

\section*{Acknowledgments}
This research was pursued under the EPSRC `Working Together: Constraint Programming and Cloud Computing' grant, a Royal Society Industry Fellowship `Bringing Science to the Cloud', and an Amazon Web Services Education Research Grant.

\bibliographystyle{ieeetr}
\bibliography{references}

\end{document}

%% file: Intro.tex
Scientists often combine highly distributed data and services through a workflow \cite{barker2007scientific}: a set of coordination rules that form a distributed application, which is executed by a workflow engine. A workflow is usually specified from the view of a single participant, and orchestrated using a single centralised engine. This means that a central engine coordinates all the services involved in a workflow, and all data flow through it. 
The location of the engine is not typically a factor, which is considered when executing a workflow. However, for collaborative scientific workflows in which the services are data-intensive and spread across multiple geographical regions \cite{Lu:ICWS:2009}, the data might have to move across long geographical distances to flow through the centralised workflow engine. This in turn degrades the overall performance of a distributed workflow \cite{Luckeneder:2013:CloudCom}.


One solution to this problem is to move the engine to an optimal location based on the geographical location of the services in a workflow. Using public cloud infrastructure such as Amazon Elastic Compute cloud (EC2)\footnote{http://aws.amazon.com/ec2/} it is possible to deploy a workflow engine automatically into a suitable region that is geographically closer to the web services in the workflow - the expectation being that the overall execution times of a workflows will be reduced. A more sophisticated approach, adopted in this paper, is a decentralised deployment in which a number of engines are deployed across the EC2 regions to orchestrate the workflow. When workflows consist of a large number of geographically distributed services it is a considerable challenge to determine how to locate the orchestration engines for optimal execution times.

\begin{figure}[h]
\subfloat[Workflow]     {\includegraphics[width = 0.15\textwidth]{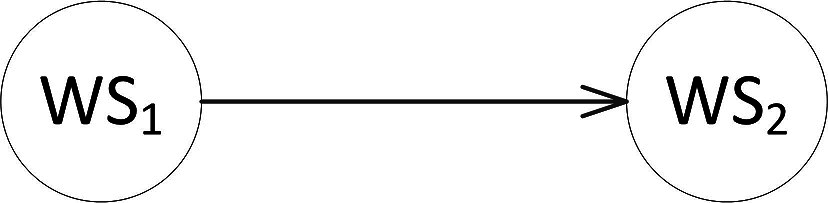}\label{fig1a}} \hfill
\subfloat[Centralised]    {\includegraphics[width = 0.15\textwidth]{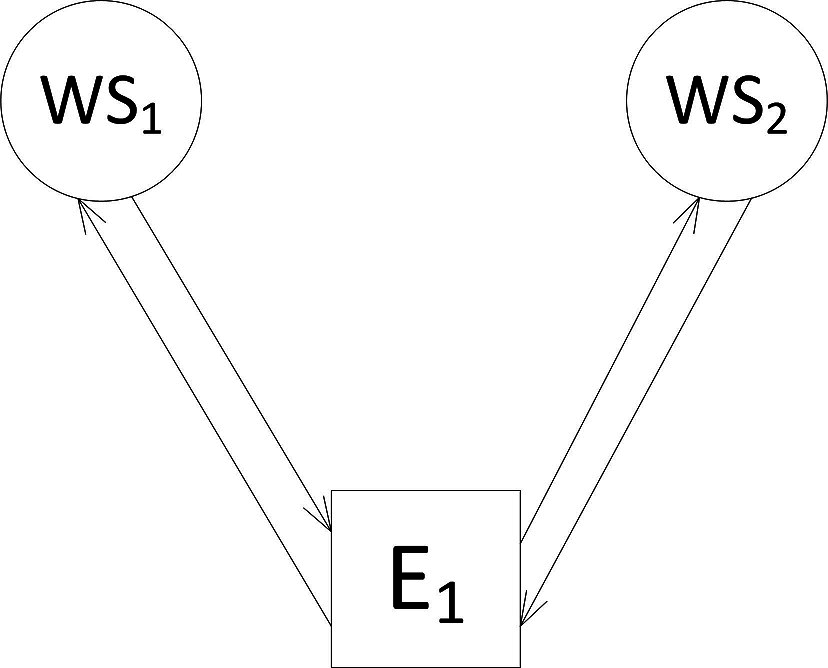}\label{fig1b}} \hfill
\subfloat[Decentralised]  {\includegraphics[width = 0.15\textwidth]{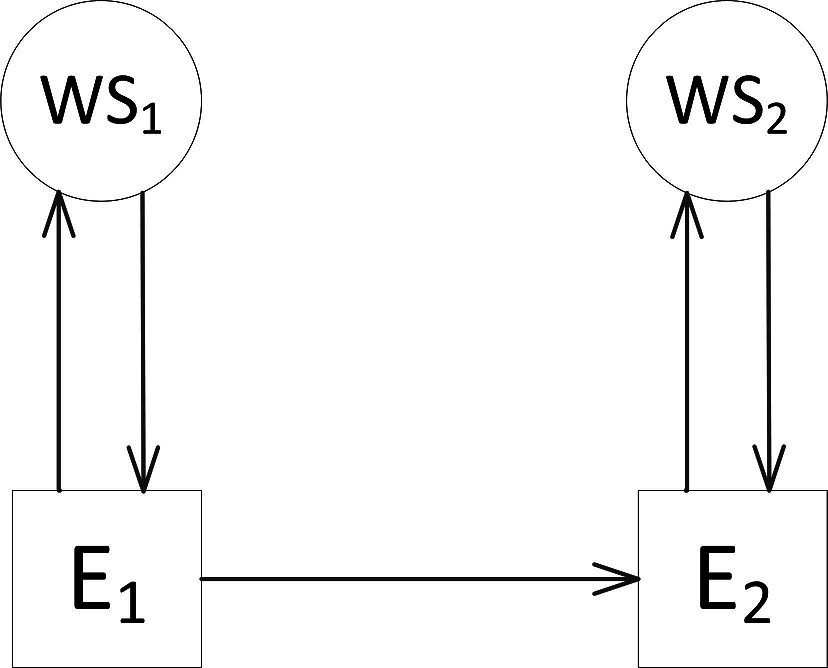}\label{fig1c}}
\caption{Workflow fragments}
\label{workflow-fragments}
\end{figure}

Consider the fragment of a workflow comprising two web services $WS_1$ and $WS_2$ shown in Figure \ref{fig1a}.
Figure \ref{fig1b} shows a centralised approach in which an engine $E_1$ invokes $WS_1$ and uses its output to invoke $WS_2$.
The location of $E_1$ with respect to the web services is crucial to the overall execution time of the workflow.
Figure \ref{fig1c} shows a decentralised approach where two engines $E_1$ and $E_2$ are employed.
Such a decentralised deployment can potentially reduce execution time by placing engines closer (in terms of network distance) to the web services.

In this paper, we investigate where the engines need to be deployed in the cloud, whether in a centralised or a decentralised fashion, so as to minimise the total expected execution time of the workflow. In addition to the workflow itself, the parameters of this problem are based on the communication costs between each web service and possible locations for the engines. The solution to the problem is the locations of engine invoking web services in the workflow.

Our research is facilitated by the development of a multi-component framework, which employs a constraint programming solver \cite{akgun_extensible_2011} to find an optimal deployment of workflow engines on the cloud. The engines support high levels of decentralisation by allowing intermediate data to be transferred between one another.

The remainder of this paper is structured as follows: Section II presents a model of the optimal deployment problem. Section III introduces the framework and the workflow engine implemented in this research. Section IV considers the experimental studies and discusses the results. Section V presents related work and Section VI concludes this paper.


%% file: Model.tex
In this section, we present a mathematical model, which captures the workflow deployment problem and how it is solved using constraint programming. 

\subsection{Problem Modelling}



Consider a set of web services and engines, which are represented as $S = \{s_1, s_2, \cdots\}$ and $E = \{e_1, e_2, \cdots\}$ respectively. For simplicity, we use $s_i \in S$ and $e_j \in E$ to represent the geographical locations of a service and an engine. The services and engines can be deployed in the same location. Let $E_u \subset E$ denote the engines used in the workflow and $|E_u| \geq 1$ is the number of used engines.

The size of a service's input is $in_{s_i}$ and of the output is $out_{s_i}$. While these do not indicate the actual size of the data, the ratio of the input and output data is captured.

A workflow needs to specify data movement between web services. For a workflow with $n$ services, we choose to denote this as a set of pairs $WF = \{(s_i, s_j), \cdots\}$, where $i \in \{1, 2, \cdots, n-1\}$, $j \in \{2, 3, \cdots, n\}$ and $i \neq j$. The former service in the pair produces data that is consumed by the latter. 
We define $p(s_i) \subset S$ as a set of web services immediately preceding $s_i$, or in other words the services that produce inputs for $s_i$. If a service $s_j \in S$ does not have any preceding services, then $p(s_j) = \emptyset$.

The cost of moving one unit of data between services and engines is represented as:
\begin{equation}
  c_{i, j} = \left\{
    \begin{array} {l l}
      0     & \quad \text{if $i = j$ and $i, j \in E$} \\
      \infty  & \quad \text{if $i, j \in S$} \\
      0 < c_{i,j} < \infty  & \quad \text{otherwise} \\
    \end{array} \right.
\end{equation}
We assume that there is no cost for communicating between the same engine as data already resides on it. The costs of moving data between two web services is infinity since they cannot communicate with each other without the mediation of an engine. Otherwise, the cost for communicating between a service and an engine or between one engine and another engine is estimated before the deployment of the workflow. 

We define $e_{s_i} \in E$ as the engine invoking a service $s_i \in S$. The cost to invoke a service is the time it takes for input data to travel from an engine to a service and for the output data from the service back to the engine. This cost is defined as:
\begin{equation}
  invoCost_{s_i} = c_{e_{s_i}, s_i} \times in_{s_i} + c_{s_i, e_{s_i}} \times out_{s_i}
\end{equation}

We define $costUpTo$ as the total data movement cost which is the sum of the cost to invoke a service and the cost to move the data required by the service to the invoking engine. A fan-in pattern (multiple web services produce inputs for a single web service), can be executed in parallel. Hence, the cost of moving data to an engine that invokes the consuming web services is represented as:

\begin{equation} \label{eq:cost_up_to}
  \begin{split}
    costUpTo_{s_i} &= \max_{s_j \in p(s_i)}{(costUpTo(s_j) + c_{e_{s_{j}}, e_{s_i}} \times out_{s_j})}\\
    &+ invoCost_{s_i}
  \end{split}
\end{equation}
The equation indicates that there is a dependency between the location of the consuming engine and the location of the producing engines. 

The total data movement time is calculated as:
\begin{equation}
  total\_movement = max_{s_{i} \in S}{(costUpTo_{s_{i}})}
\end{equation}

In most cases $total\_movement$ is the value of $costUpTo$ of the last service in the workflow. However, if there are multiple independent services at the end of the workflow, we need to select the largest $costUpTo$ value among them as the $total\_movement$.

We define $costEngineOverhead$ as the penalty for adding any additional engines in the workflow. This value can be used to limit the number of engines that a user requires; adding more engines can increase costs. Hence, the total overhead in the workflow is:
\begin{equation}
  total\_overhead = costEngineOverhead * (|E_u| - 1)
\end{equation}
Now the total cost of executing a workflow is
\begin{equation} \label{eq:total_cost}
  total\_cost = total\_movement + total\_overhead
\end{equation}

The optimal deployment plan that needs to be generated is 
the mapping between the services and the engines (the set of $e_{s_i}$ for $s_i \in S$) so that $total\_cost$ can be minimised.

\subsection{Solving the Problem Using Constraint Programming}


In order to solve the above problem, we first need to represent the mathematical model in a suitable format for a Constraint Programming (CP) solver.
Producing a correct and efficient CP model for a problem like this is a challenging task.
Rather than crafting a constraint model by hand, we employ the automated constraint modelling system \conjure{} \cite{akgun_extensible_2011}.
In order to do so, we specify the problem in \essence{} \cite{EssenceConstraints} which is a high level problem specification language which offers abstract mathematical constructs and a rich collection of operators, i.e. users are not required to make the large number of low level modelling decisions they would otherwise need to make.

%% file: Framework.tex
\begin{figure}[h]
	\vspace{-5mm}
		\includegraphics[width = 0.45\textwidth]{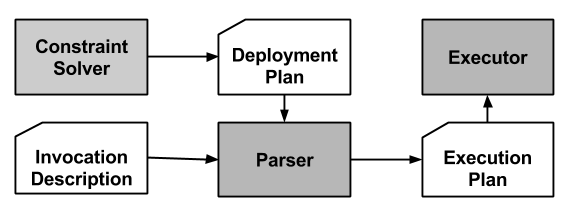}
	\caption{\label{framework}Framework Architecture}
\end{figure}

As shown in Figure \ref{framework} the framework has three components, namely the Constraint Solver, the Parser and the Executor, and three script files, namely the Invocation Description, the Deployment Plan and the Execution Plan. Script files are chosen as the data transfer mechanism since they facilitate reproducibility for future experiments without running the whole process again, and interoperability between the components.

The Constraint Solver uses the workflow $WF$, the inputs and outputs ($in_{s}$ and $out_{s}$, where $s \in S$) and the cost $c$ between web services and engines to produce the Deployment Plan. This output is then used by the Parser along with the Invocation Description to generate an Execution Plan, which is then executed by the Executor. In order to demonstrate the process in which a execution plan is produced for a given workflow, we focus on explaining the script files instead of the components.

\subsection{Invocation Description}

An \textbf{invocation description} describes the data flow between services and how each service is invoked. Since we use RESTful web services which are not described by a common interface, like WSDL for SOAP services, the names of all web services' parameters must be specified. Each line represents one service invocation including service name (e.g. URL), one or more inputs to the service (each input is represented as a pair of a parameter name and a corresponding value) and the output. The input parameter and value by default are strings referring to the actual values stored inside the engine. This can also be passed-by-value by wrapping the  parameter or value inside single quotes ('). The output value is a reference to the memory of the engine.

\begin{figure}[h]
	\begin{lstlisting}
ws_1 'param_1':'0' value_2
ws_2 'param_2':value_2 value_3
	\end{lstlisting}
	\caption{Invocation Description}
	\label{invoc_file}
\end{figure}

Figure \ref{invoc_file} presents an invocation description of the workflow which is presented in Figure \ref{fig1a} and consists of two web services \texttt{ws\_1} and \texttt{ws\_2}, each of which requires one parameter named \texttt{param\_1} and \texttt{param\_2} respectively. A zero value is passed to \texttt{ws\_1} whose output is stored and referenced by the key \texttt{value\_2}. The data referenced by \texttt{value\_2} is the input for \texttt{ws\_2}. The final result produced by \texttt{ws\_2} is stored in \texttt{value\_3}.

\subsection{Deployment Plan}
Based on the inputs to the Constraint Solver, which are $WF$, $in$, $out$ and the cost $c$, a \textbf{deployment plan} is produced, it is the mapping between the web services and the cloud regions in which the engines are deployed. The cloud regions and web services have a one to many relationship; one region can have many web services, but a web service can only be assigned to one region.

For example, Figure \ref{deploy_plan} shows that \texttt{ws\_1} and \texttt{ws\_2} are mapped to \texttt{region\_1} and \texttt{region\_2} respectively. As a result, \texttt{ws\_1} will be invoked by an engine in \texttt{region\_1}.

\begin{figure}
	\begin{lstlisting}
ws_1 --> region_1
ws_2 --> region_2
	\end{lstlisting}
	\caption{Deployment Plan}
	\label{deploy_plan}
\end{figure}

\subsection{Execution Plan}

In order to execute a workflow, based on the invocation description and the deployment plan, an \textbf{execution plan} is created; an example is shown in Figure \ref{exec_plan}. It describes the service invocations performed by each engine and additional steps to move data between them. Moreover, it contains information required to deploy the engines on the cloud.

\begin{figure}[h]
	\begin{lstlisting}
 1  # define hosts
 2  host region_1 aws ubuntu region_1_ip
 3  host region_2 aws ubuntu _
 4 
 5  # define engines
 6  serv eng_1 engine
 7  serv eng_2 engine
 8
 9  # deploy engines on hosts
10  depl eng_1 region_1
11  depl eng_2 region_2
12
13  # invocations for engine_1
14  eng_1 ws_1 'param_1':'0' value_2
15  eng_1 eng_2.Setter 'value_2':value2 ack_1
16
17  # invocation for engine_2
18  eng_2 ws_2 'param_2':value_2 value_3
	\end{lstlisting}
	\caption{Execution Plan}
	\label{exec_plan}
\end{figure}


Lines 2 and 3 describe the cloud regions in which the engines are deployed and includes the data required for the deployment such as cloud provider (e.g. $aws$), the username to access the remote machine (e.g. $ubuntu$) and the ip address or host name of the instance (e.g. $region\_1\_ip$). In Line 3, the ip address of $region\_2$ is $\_$ to denote that it is currently unknown, i.e. the instance is not running. When the execution begins the framework will start the cloud VM and replace $\_$ with the actual ip address. Engine definitions are presented in lines 6 and 7. \texttt{engine} refers to an actual application which will be deployed on the cloud. \texttt{eng\_1} and \texttt{eng\_2} are aliases given to the engines deployed in the cloud regions. Lines 10 and 11 describe the deployment of engines to hosts. Lines 14 and 18 represent the service invocations, which are quite similar to Figure \ref{invoc_file}. The only difference is that the engine alias is added at the beginning of each line to denote an engine performing an invocation. Line 15 is an additional step to move the output of \texttt{ws\_1} from \texttt{eng\_1} to \texttt{eng\_2}.

\subsection{Workflow Engine}


There are many existing workflow engines such as DAGMan\footnote{http://research.cs.wisc.edu/htcondor/dagman/dagman.html} or Taverna\footnote{http://www.taverna.org.uk/}. However, they have many complex features and are difficult to set up on remote machines. Hence, we decided to implement a light-weight engine which can be easily deployed on any remote machines

Our engine is a RESTful web service which can invoke other RESTful web services. It can be easily deployed on a remote machine by copying and executing its source code. Data transfer between engines is performed as a normal service invocation which contains transferred data as its input, e.g. line 15 of Figure \ref{exec_plan}.
To support parallelism, for every successful invocation, the engine finds other invocations whose all input data is available and invokes them.

%% file: Experiment.tex
\subsection{Experiment Setup}

\begin{figure}
	\subfloat[Workflow 1]{\includegraphics[width = 0.20\textwidth]{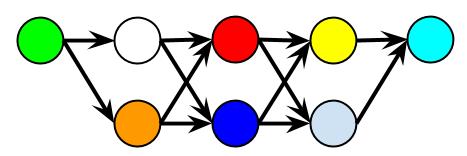}} \hfill
	\subfloat[Workflow 2]{\includegraphics[width = 0.20\textwidth]{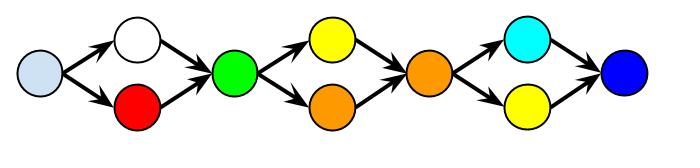}} \\
	\subfloat[Workflow 3]{\includegraphics[width = 0.20\textwidth]{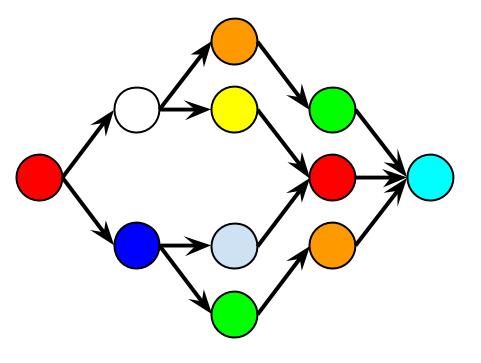}} \hfill
	\subfloat[Workflow 4]{\label{wf_10}\includegraphics[width = 0.20\textwidth]{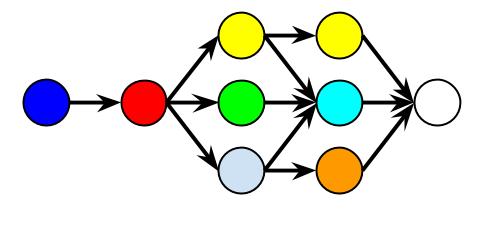}} \\
	\begin{center}
		\subfloat{\includegraphics[width = 0.45\textwidth]{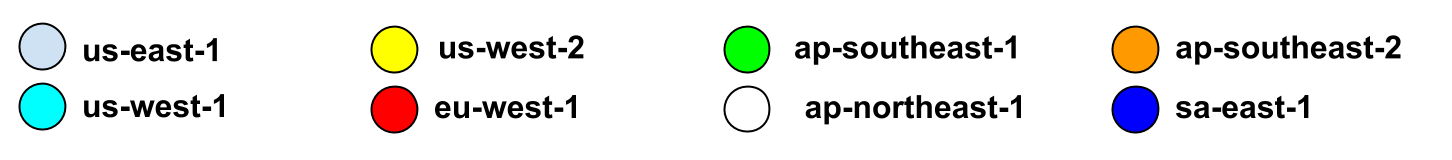}}
	\end{center}
	\caption{\label{sampleworkflows}Four sample workflows}
\end{figure}

In order to investigate the performance of our model and framework, we performed experiments on four sample workflows. These comprise between eight and eleven web services, which we have deployed across all eight EC2 regions as summarised in Figure \ref{sampleworkflows}, in which a colour of a node (i.e. service) represents its location. 
We focus primarily on Directed Acyclic Graphs (DAG) based workflows since these are heavily used in the scientific community.
We also select eight available AWS regions as potential locations to deploy the workflow engines.

Our workflows were generated based on a combination of three generic patterns found in all Directed Acyclic Graph (DAG) based scientific workflows: linear (sequence), fan-in (multiple sources mapped to one sink) and fan-out (one source mapped to multiple sinks). The generated workflows are realistic because scientists usually have no choice over the ordering of third-party web services as they must be used in a certain order to execute an end-to-end distributed application. Moreover, the workflow services are normally left unmaintained after the execution is completed.



Prior to our experiment, the mean Round-Trip Time (RTT) between the regions was measured and used as the costs of moving data for the constraint model. For each workflow, we used the CP model and the $costEngineOverhead$ value in order to have multiple solutions using different numbers of engines. In other words, we tried to execute the workflow using from 1 to multiple engines.

For comparison, we measured the execution time of these workflows using two naive (yet realistic) approaches: a single orchestration engine running at the user's host (in our case St Andrews), and a single orchestration engine running on the nearest EC2 region (in our case, Dublin).

\subsection{Results and Discussion}

\begin{figure*}
\vspace{-5mm}
\subfloat[Workflow 1 ]{\includegraphics[width = 0.25\textwidth,height=0.18\textheight]{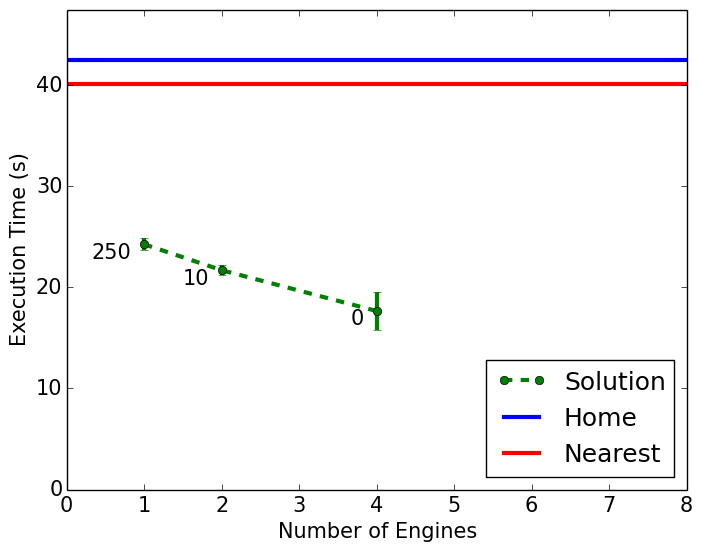}} \hfill
\subfloat[Workflow 2 ]{\includegraphics[width = 0.25\textwidth,height=0.18\textheight]{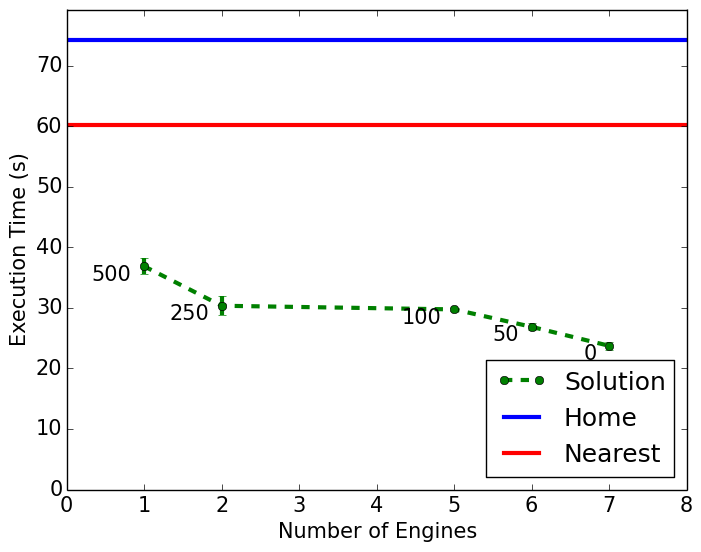}} \hfill
\subfloat[Workflow 3 ]{\includegraphics[width = 0.25\textwidth,height=0.18\textheight]{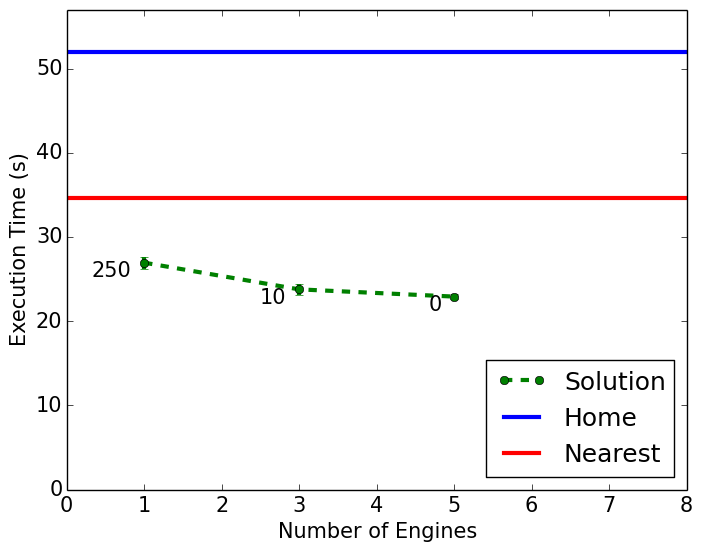}} \hfill
\subfloat[Workflow 4 ]{\includegraphics[width = 0.25\textwidth,height=0.18\textheight]{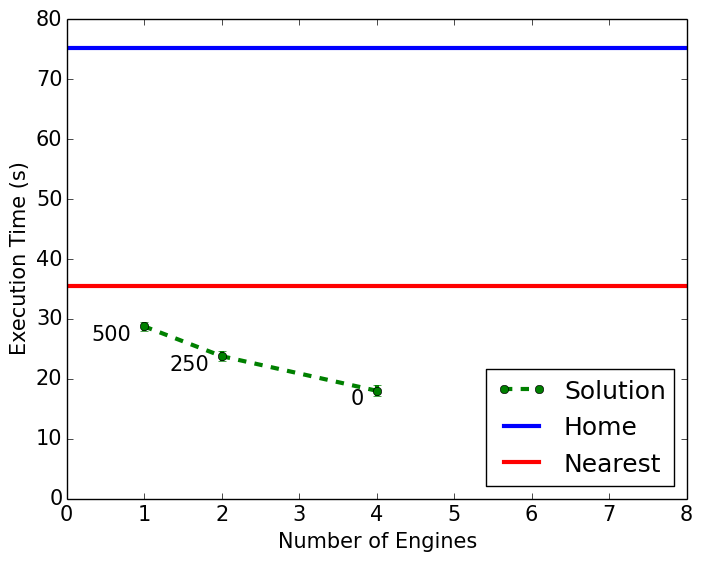}}
\caption{\label{sampleworkflows-times}Execution times for the workflows}
\end{figure*}

Figure \ref{sampleworkflows-times} displays the results of our experiment. The x-axis shows the number of engines while the y-axis presents the total execution time.

The naive solutions are presented by blue (and red), when the engine was deployed in St Andrews (and Dublin), thick lines. Since we assumed that the communication between engines at the same location is instantaneous and the naive deployments are centralised, they are therefore not affected by the number of engines, hence, a straight line plot.

The green dashed line represents the execution time of our solutions. Each deployment was executed 15 times. Each point in the plots represents the mean of these executions (excluding the slowest 5 executions to account for network instability) and the error bars show the standard deviation. The number at each point represents the $costEngineOverhead$ value.

It is immediately evident that the solutions provided by our framework performed considerably and consistently better than the naive single-orchestrator approaches. Even if only one engine is allowed our framework can produce a solution with better performance. For all workflows, the solution with more engines always had the better execution time; using more engines reduced the cost of moving data between web services. Moreover, none of the workflows used all of 8 possible locations as the optimal solution. In other words, completely decentralising the workflow, i.e. assigning each service to an engine deployed at the same host as it, does not guarantee the best performance.

\begin{figure}
	\begin{tabularx} {0.45\textwidth} {| X | X | X | X | X | X | X | X | X | X | X |}
		\hline
		Workflow & 1 & 2 & 3 & 4  \\
		\hline
		Minimum & 1.7 & 1.6 & 1.6 & 1.3 \\
		\hline
		Maximum & 2.3 & 2.5 & 1.5 & 2.0 \\
		\hline
	\end{tabularx}
	\caption{Minimum and Maximum speedup when compared to nearest execution}
	\label{speedup}
\end{figure}

In order to compare the improvement between using our framework and the naive solutions, we calculated the speedup between the plan in which a centralised engine is deployed at Dublin, the least optimal (one engine) and most optimal (maximum number of engine possible). The results are presented in Figure \ref{speedup}, which shows that our framework is able to improve the performance from 1.3 to 2.5 times.

\begin{figure*}
	\vspace{-5mm}
	\subfloat[One Engine]{\label{wf_10_1}\includegraphics[width = 0.3\textwidth,height=0.1\textheight]{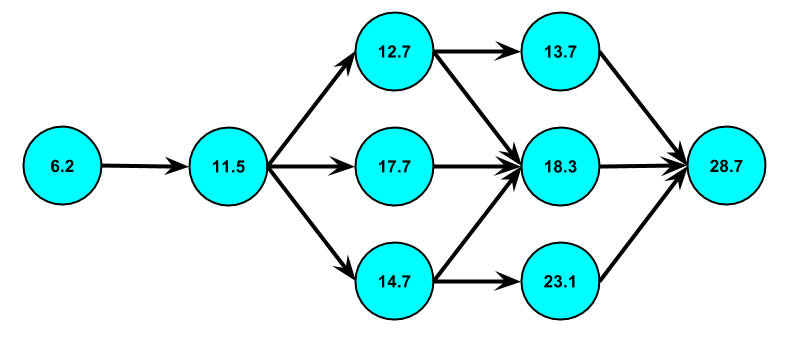}} \hfill
	\subfloat[Two Engines]{\label{wf_10_2}\includegraphics[width = 0.3\textwidth,height=0.1\textheight]{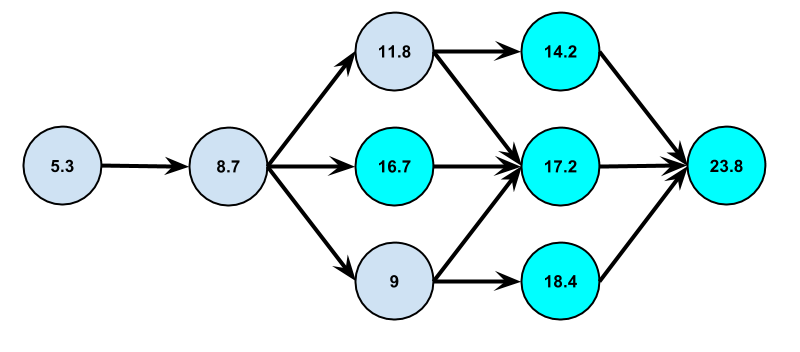}} \hfill
	\subfloat[Four Engines]{\label{wf_10_4}\includegraphics[width = 0.3\textwidth,height=0.1\textheight]{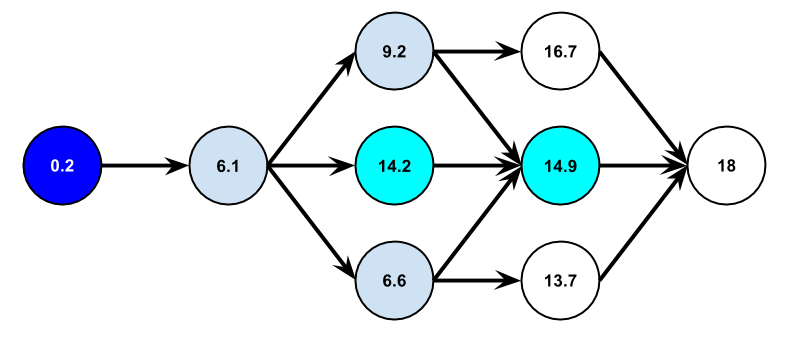}}
	\begin{center}
		\subfloat{\includegraphics[width = 0.35\textwidth]{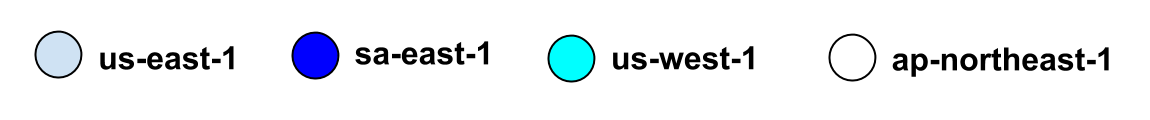}}
	\end{center}
	\caption{\label{wf_exec_plan}Execution Plans for Workflow 4}
\end{figure*}

Figure \ref{wf_exec_plan} presents the execution plans using 1, 2 and 4 engines, respectively, for the workflow number 4 in Figure \ref{sampleworkflows}. Notably, the colour does not represent the location of a service but the location of an engine invoking it. The number inside each node (i.e service) is the number of seconds it takes to finish invoking that service after starting executing the workflow, i.e. the actual $costUpTo$ value. Which also means that the number in the last node is the total execution time of the workflow. Figure \ref{wf_exec_plan} shows that by using more engines, the workflow is partitioned into smaller and dependent sub-workflows. The result of this partitioning is the lower $costUpTo$ at most of the services. Notably, the goal of the CP model is to minimise the $costUpTo$ of the \textbf{last} service in the workflow, not all of them. Hence, even though there is an service in \ref{wf_10_4} with higher $costUpTo$ values than the same ones in \ref{wf_10_2}, its total execution time (i.e. the $costUpTo$ of its last service) is still lower.

%% file: LiteratureReview.tex
\subsubsection{Decentralised Workflow Orchestration}

In \cite{Javadi:2012:CCGrid}, the authors decentralised a workflow by modifying the services to transfer data between services. Instead of directly changing the services, the authors of \cite{Binder:2006:ICWS} proposed an additional layer which stored and triggered invocation if all data was available. However, these studies cannot be easily applied when services are managed by external organisations and do not allow them to be modified. In our previous work \cite{Barker:2008:EMP}, we proposed a framework in which a proxy was assigned to one or more services and invoked them based on instructions given by a centralised engine. By deploying proxies near their services, it reduced the data transfer overhead. Multiple proxies mitigate the bottleneck caused by using a single centralised controller. In this paper, instead of one engine controlling multiple proxies, we used multiple engines without proxies, each of which executes a sub-workflow. Moreover, we aimed to find the optimal locations to deploy engines.

\subsubsection{Workflow Partitioning}

Workflow partitioning aims to split a workflow to smaller fragments, each of which is executed by an engine. Pegasus \cite{Chen:2011:PSW} has many advanced mechanisms for partitioning and resource mapping. However, it does not consider the geographical distribution of services. Similarly, in \cite{Chen:2010:PDCAT}, the services are grouped based on the functional similarity between them and each group is assigned one engine. The research aimed at reducing the communication traffic between web services and engines without taking into account the network distance. Hence, it is unclear if this approach can reduce total execution time.



\subsubsection{Data and Location Aware Workflow Execution}

The authors of \cite{Catalyurek:2011:IDP} proposed approaches to dynamically re-locate data and task in order to achieve data locality/proximity. However, if data and services are managed by external organisations they cannot be re-located and these approaches may not be applicable. Our previous work \cite{Luckeneder:2013:CloudCom} aimed to find an optimal location to deploy a centralised engine based on both geographical and network distances. In this paper, we further expand our research by decentralising a workflow while taking into account its geographic distribution. We also consider the parallelism of workflow execution instead of assuming that the execution is sequential.

%% file: Conclusion.tex
This paper discussed how to improve the performance of highly distributed workflows by decentralising the orchestration and selecting cloud locations to deploy workflow engines. The workflow deployment problem is modelled and solved using constraint programming in which a constraint solver produces the optimal deployment plan. 


By comparing our approach to traditional ones in which the centralised workflow engine was deployed at either our home location (St Andrews, Scotland) or the nearest AWS region (Dublin, Ireland), it was evident that our approach was able to reduce the data movement cost, and thus resulted in better execution time. Our approach demonstrated that RTT is a reliable metric to calculate network distance.

We developed a framework for our experiments which generates an optimal plan for executing workflows, manages cloud VMs and executes workflows. A lightweight workflow engine was also developed, which can be easily deployed on any remote machine and is able to perform multiple independent service invocations concurrently.

In the future, we aim to estimate the actual time for web service invocations to schedule workflow deployment during runtime instead of starting all VMs ahead of time. We also plan to develop a dynamic monitoring and planning mechanism to adapt to network changes during the execution.